\documentclass[12pt]{article}
\usepackage[utf8]{inputenc}
\usepackage{Donnay_Preamble}
\usepackage{geometry}
\newcommand{\blind}{0}

\def\spacingset#1{\renewcommand{\baselinestretch}%
{#1}\small\normalsize} \spacingset{1}

\title{3:1 Nesting Rules in Redistricting}
\author{Christopher Donnay\thanks{Department of Mathematics, The Ohio State University}}
\date{\today}

\begin{document}
\if0\blind
{
  \title{\bf 3:1 Nesting Rules in Redistricting}
  \author{Christopher Donnay\thanks{
    This material is based upon work supported by the National Science Foundation under Grant No. DMS-1928930 and by the Alfred P. Sloan Foundation under grant G-2021-16778, while the first author was in residence at the Simons Laufer Mathematical Sciences Institute (formerly MSRI) in Berkeley, California, during the Fall 2023 semester.
}\hspace{.2cm}\\
    Department of Mathematics, The Ohio State University}
  \maketitle
} \fi

\if1\blind
{
  \bigskip
  \bigskip
  \bigskip
  \begin{center}
    {\LARGE\bf Title}
\end{center}
  \medskip
} \fi

\bigskip
\begin{abstract}
In legislative redistricting, most states draw their House and Senate maps separately. 
Ohio and Wisconsin require that their Senate districts be made with a 3:1 nesting rule, i.e., out of triplets of adjacent House districts.
We seek to study the impact of this requirement on redistricting, specifically on the number of seats won by a particular political party. 
We compare two ensembles generated using Markov Chain Monte Carlo methods; one which uses the ReCom chain to generate Senate maps without a nesting requirement, and the other which uses a chain that generates Senate maps with a 3:1 nesting requirement.
We find that requiring a 3:1 nesting rule has minimal impact on the distribution of seats won.
Moreover, we study the impact the chosen House map has on the distribution of nested Senate maps, and find that an extreme seat bias at the House level does not significantly impact the distribution of seats won at the Senate level.
\end{abstract}

\noindent%
\vfill

\newpage

With the exception of Nebraska,  each state in the U.S. has a bicameral legislature, meaning there is an upper and lower house \cite{ballotpedia}.
Most states draw district maps for their two houses separately.
Eight states require that the state Senate map be made of pairs of adjacent districts from the lower house; we call this a \emph{2:1 nesting rule}.
Two states, Ohio and Wisconsin, require a \emph{3:1 nesting rule}, in which the state Senate map is made of triplets of adjacent districts from the lower house.

\begin{figure}[h]
\centering
\includegraphics[scale=.25]{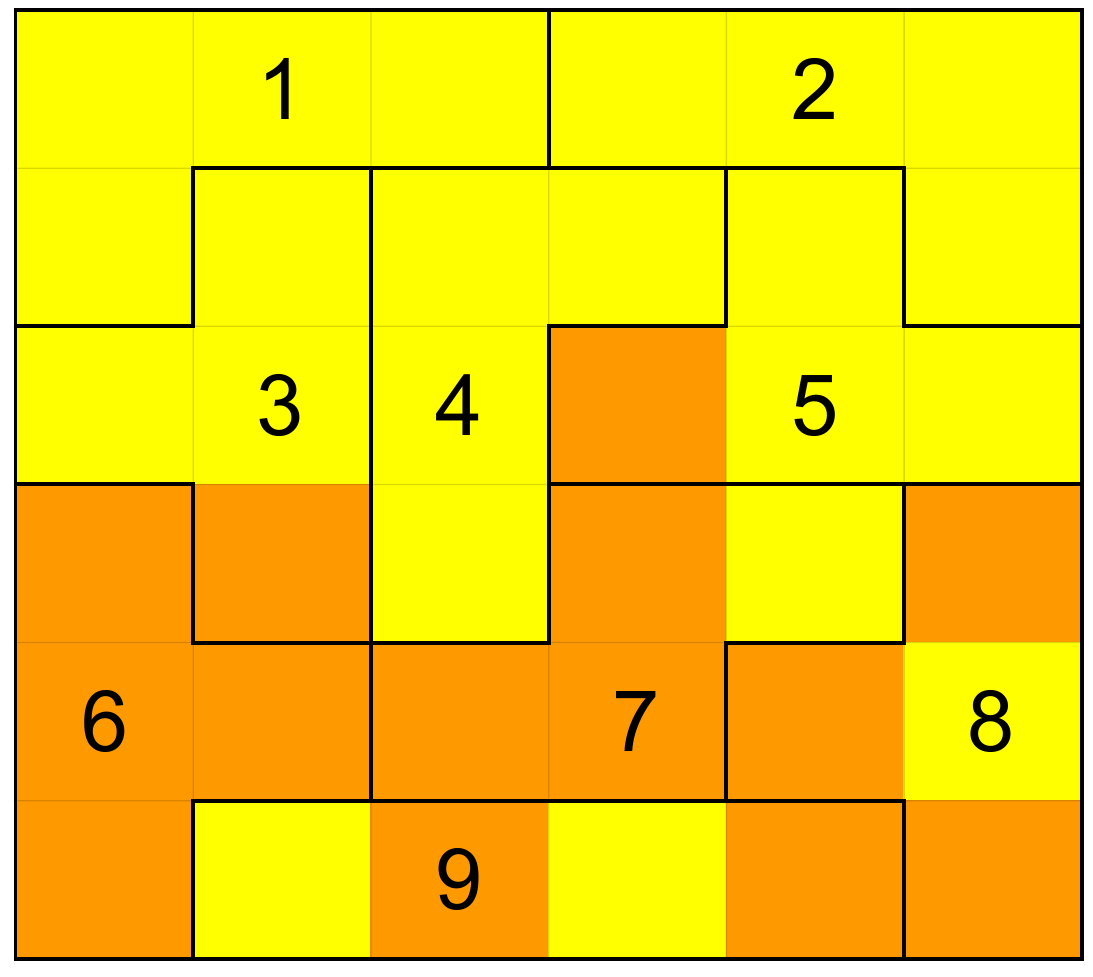}\includegraphics[scale=.2]{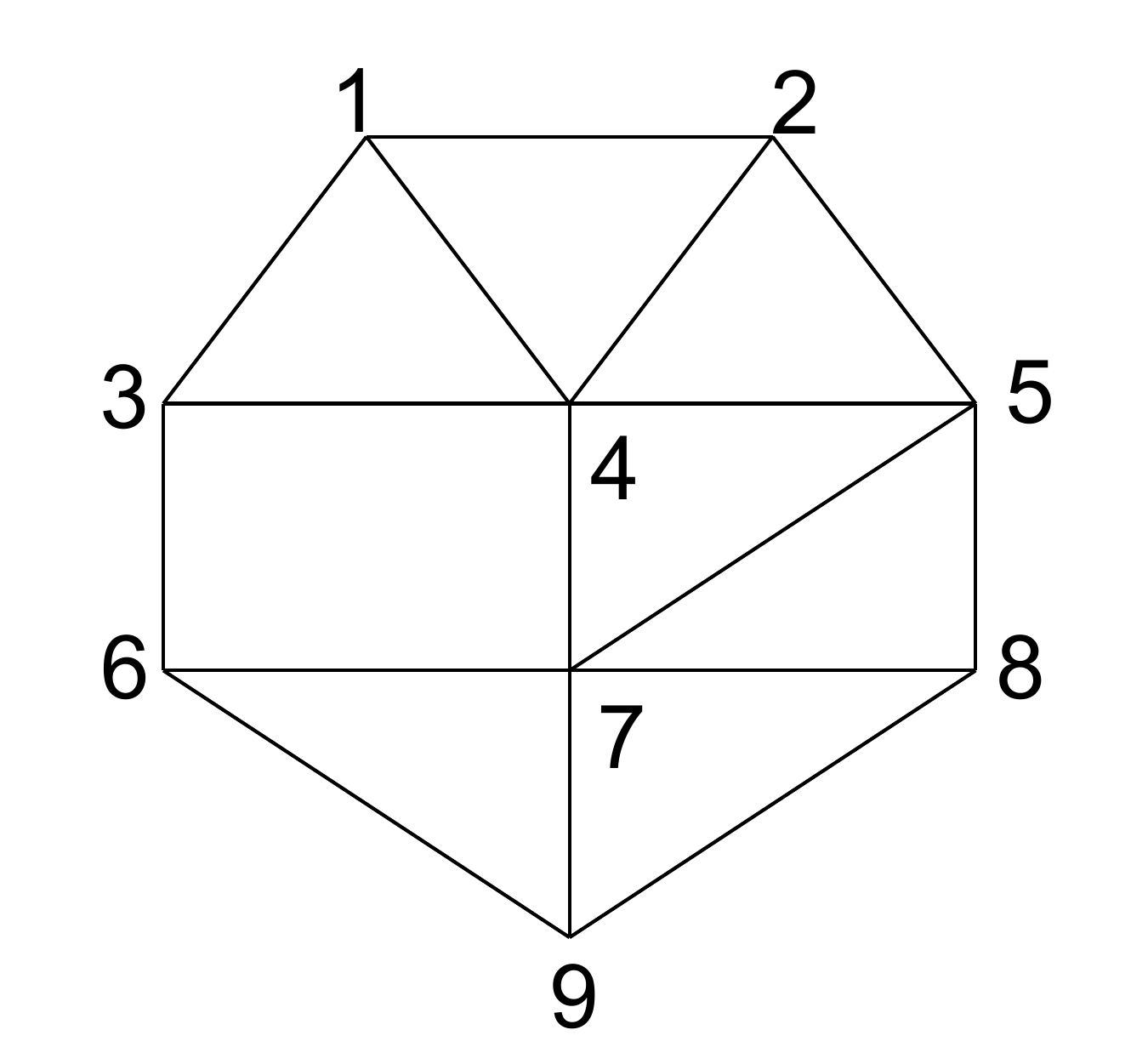}
\caption{The $6\times 6$ grid, where each square represents one voter, either of the yellow or orange party. We have chosen a fixed partition into nine House districts and constructed the corresponding dual graph.}
\label{fig: toy_example}
\end{figure}

In theory,  these nesting rules severely restrict the number of possible maps that can be drawn.
Consider the toy example on the $6\times 6$ grid in Figure \ref{fig: toy_example} from \cite{durham_2020}.
There are 264,500 ways to construct a Senate plan with 3 districts with no restrictions \cite{mggg_grid_partitions}.
However, if we fix a lower house map with nine districts, and only consider Senate maps made out of triplets of adjacent districts, we find there are only 14 Senate plans.
Common wisdom says that this restriction on the space of redistricting plans should reduce the variability of the plans.
But even in this toy example, we see that under the matching $\{(1,2,4),(3,6,9),(5,7,8)\}$,  the yellow party wins one seat, while in the matching $\{(1,3,6),(2,5,8),(4,7,9)\}$, yellow wins all three seats.
This suggests that even with a restricted state space,  a significant variation in election outcome is possible.

In this paper,  we will analyze the impact of these 3:1 nesting rules on redistricting.
The general strategy will be to take a fixed House map,  generate a large sample of possible Senate maps that follow the 3:1 nesting rules, and analyze the behavior of these maps.
Sampling from the space of 3:1 nested maps will be the only feasible choice, as counting or constructing all such maps is computationally intractable.
We will use Markov Chain Monte Carlo (MCMC) methods to sample both 3:1 nested Senate maps and Senate maps that have no such restriction.
We will also explore how the choice of initial House map impacts the distribution of possible Senate maps.

\section{The Mathematics of Redistricting}
\subsection{Dual Graphs}
Redistricting is a discrete problem.
States are divided into small units,  like precincts or Census blocks,  and these discrete units are then assigned to districts.
Mathematically, we formulate this problem using a \emph{dual graph}.
The vertices of the dual graph $G$ are the discrete units, and an edge denotes geographic adjacency, where we require that two units share a border of positive length.
A \emph{districting plan} $D = (D_1,\dots, D_n)$ is a partition of the vertices of the dual graph into $n$ components, i.e., districts.
By both federal and state law, districting plans must meet a host of other requirements.
Some of the most typical ones include
\begin{enumerate}
\item contiguity; each component of the partition must form a connected induced subgraph of $G$. 
That is, districts must be connected.

\item population balance; since the U.S.  Supreme Court ruling in Reynolds v.  Sims in 1964 \cite{ReynoldsVSims},  districts must have roughly equal population.  We usually take this to be within 5\% of the ideal population, but map drawers usually can balance this to within just a few individuals \cite{NCSLRedistricting}.

\item compactness; this is hard to operationalize, but people do not like seeing ``snakey" districts.  
\end{enumerate}

We take our dual graph to be the \emph{House dual graph}, in which the vertices represent House districts. 
A \emph{3:1 nested Senate map} is thus a districting plan on the House dual graph with the further constraints that each component of the partition be of size 3 and each component be connected.
We do not take population balance into account; we assume that the underlying House dual graph is already population balanced.
When making nested maps, we assume that the chosen House map is compact, and do not impose further compactness restrictions on the Senate maps.

\subsection{Ensemble Analysis}
\emph{Gerrymandering} is the process of drawing districts to advantage one class of people over another.
The question of how to tell if a map is gerrymandered goes well beyond the scope of this paper and crosses numerous academic disciplines and practical issues.
One prominent method in the literature is the use of \emph{ensemble methods} (see Chapter 16 and 17 of \cite{duchin_walch_2022}).
Using some generative process, a large number, or \emph{ensemble}, of districting plans are constructed for a particular state.
Then,  a proposed or enacted map can be compared to the ensemble, and if it is an outlier in some statistic,  that might indicated that it is gerrymandered.

One generative method used in the redistricting literature for sampling graph partitions is to use Markov Chain Monte Carlo (MCMC) methods (see, for example, \cite{recom_markov}) . 
Informally,  a Markov chain is a process that takes an initial object,  updates that object via some probabilistic rule,  and returns the new object.
This is one step of the chain; MCMC methods involve taking many steps in the chain, and using the generated objects as a sample.
The ReCom Markov chain was first introduced in \cite{recom_markov} to generate districting plans that meet the generally accepted standards for ``good" districts.
Given a dual graph and an initial districting plan,  the ReCom chain merges two adjacent districts, generates a spanning tree on the resulting induced subgraph, and randomly cuts the spanning tree, resulting in two new districts.
In this paper, we use the ReCom algorithm on the precinct/ward dual graphs for Ohio/Wisconsin to generate ensembles of Senate plans that do not follow any 3:1 nesting rule.
That is, we ask ReCom to generate plans with 33 districts using precincts/wards as building blocks.

ReCom does not care about the number of precincts/wards/basic units within its districts.
Its acceptance of maps is based on compactness, contiguity, and population balance.
Thus, in order to sample 3:1 nested maps, we require a different Markov chain that ensures that each component of the partition has exactly 3 vertices in it. 
It turns out that the new chain, which we call the Swap chain, is equivalent to running ReCom on the House dual graph if we assume each House district has population 1 and ask for perfect population balance. 
However, the spanning tree step of ReCom is more computationally intensive than our implementation of the Swap chain.
See Section \ref{section: swap chain} for a rigorous definition of Swap.

\section{Prior Work}
Constructing maps using nesting rules has  been proposed as a game-theoretic way of generating fair maps \cite{palmer_schneer_deluca_2022}.
In  \cite{nested_districts}, the authors explore how 2:1 nesting rules impact redistricting.
They found that in Alaska,  the 2:1 nesting rule had minimal impact on how many seats Democrats could win.
In other words, they could win the same range of seats with or without the 2:1 nesting rule.

In \cite{nested_districts}, the authors are able to do two kinds of analysis.
When computationally feasible, they construct all possible 2:1 nestings on a House dual graph, and study the properties of the maps.
When it is infeasible to construct all 2:1 nestings,  they can uniformly sample them thanks to a connection to perfect matchings and the FKT algorithm.
Unfortunately,  constructing all possible 3:1 nested plans for Ohio or Wisconsin is computationally infeasible.
The 3:1 nesting requirement can be formulated as a problem about perfect matchings on 3-uniform hypergraphs, which is one of Karp's original NP-complete problems \cite{Karp1972}.
Moreover,  the method of uniform sampling used in \cite{nested_districts} does not extend to 3:1 nestings.

\subsection{Swap Chain}
\label{section: swap chain}
\begin{figure}[h]
\center
\includegraphics[scale=.5]{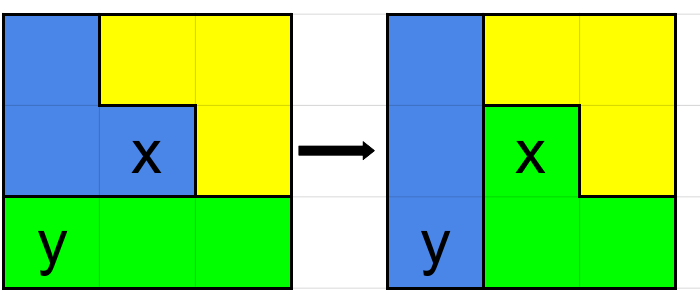}
\includegraphics[scale=.5]{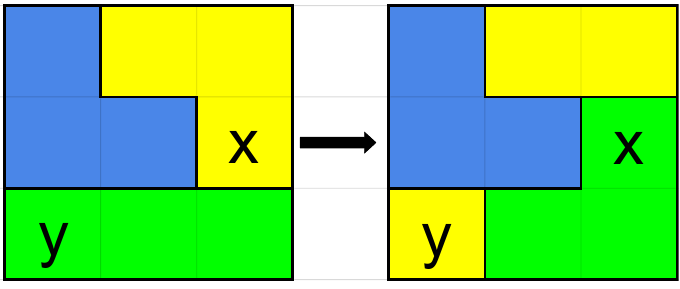}
\caption{The left side  shows a valid swap of two House districts, while the right side shows an invalid swap, since it disconnects the yellow district.}
\label{fig: swap chain grid validity}
\end{figure}

In order to sample 3:1 nested Senate maps, we make use of a Markov chain first introduced by Durham in \cite{durham_2020},  which we will call the \emph{Swap} chain.
Given a 3:1 nested Senate map, to take a step in the chain, proceed as follows.
\begin{enumerate}
\item Choose two House districts uniformly at random with replacement.

\item Swap the Senate district assignment of the two House districts.

\item If this swap does not create a valid (i.e., contiguous) Senate map,  go back to step 1.

\item If the swap does create a valid (contiguous) Senate map,  accept the new map.
\end{enumerate}
See Figure \ref{fig: swap chain grid validity}.

A recent arXiv preprint shows Swap is not always irreducible  \cite{tuckerfoltz2023locked}.
That is, we cannot always get from one districting plan to every other one via Swap moves.
However, this theoretical result is about grid graphs, and we see no evidence that our chain is not effectively moving through the state space for real world dual graphs.

\section{Our Data}
In order to analyze maps produced by our Markov chains, we need to know how people vote.
Our best substitute for this is to take past election data and assume that people living in a district will vote the way they did in a previous election.
Unfortunately, voting data is reported at the precinct/ward group level in Ohio and Wisconsin,  while districts are composed of Census blocks.
We thus rely on third party collection sources who put in an enormous amount of effort creating digital SHP files that record the boundaries of precincts/wards along with election data.
In this paper, we use both the MGGG Redistricting Lab's (MGGG) and the Redistricting Data Hub's (RDH) collection of SHP files that include precinct boundaries and election data.
For more information on which SHP files we used,  how we included population data,  any preprocessing decisions made before running our Markov chains,  and our code for running the chains,  see \href{https://github.com/cdonnay/nesting_OH_WI}{our GitHub repository} \cite{cdonnayNesting}.

In Ohio, out of the nine statewide elections for which we had precinct files, we selected the 2018 Senate (SEN18) and Treasurer (TRES18) race as our election data.
The 2018 Senate and Treasurer election data have a very similar vote percentage, but flipped for each party.
In SEN18 we have 53.4\% for the Democratic candidate and 46.6\% for the Republican, while in the TRES18 we have 53.5\% for the Republican and 46.7\% for the Democrat.
There is not a major third party/write-in presence in either race.

In Wisconsin, out of the six statewide elections since 2016 for which we had SHP files,  we decided to work with the 2018 Senate (SEN18) and Attorney General (AG18) races.
The 2018 Senate race has the lowest Republican vote share out of all of the available races (44.6\%), while the 2018 Attorney General race has the highest (48.8\%).

\section{Exploring Convergence }
It is important to ensure that we have taken enough steps of our Markov chain to converge to the underlying distribution.
To explore the convergence of our chains, we proceed as in \cite{colorado_in_context} and \cite{2018ComparisonOD}; we choose relevant summary statistics for our ensembles, project to said statistics, and explore convergence there.
Of course, none of this precludes pseudo-convergence.
We choose the number of Democrat seats won as our statistic.
We use several heuristics to explore convergence:
\begin{enumerate}
\item We compute the $n$-lag autocorrelation of the desired statistic, i.e.,  the Pearson correlation between the series and itself shifted by $n$.  An autocorrelation of lag $n$ close to 0 indicates no linear correlation between the statistic at one time and $n$ steps in the future. 
Autocorrelations that quickly decay to 0 indicate a fast-mixing Markov chain \cite{roy2019convergence}.

\item We examine the first 10\%, 50\%, and 100\% of the run of the chain, and see if our desired statistic has stabilized.  

\item We examine runs of the chain started from different seeds, and see if the desired statistic is reasonably similar across seeds.
\end{enumerate}

We also note that we will not show each figure with every type of election data or for each type of chain.
However,  all of our results are consistent across such choices unless otherwise noted, and the full set of figures can be found in our GitHub repository \cite{cdonnayNesting}.
As a result of our discussion below,  in our final data runs, we ran our Swap and ReCom chains for 1 million steps.

\begin{figure}
\center
\includegraphics[scale=.5]{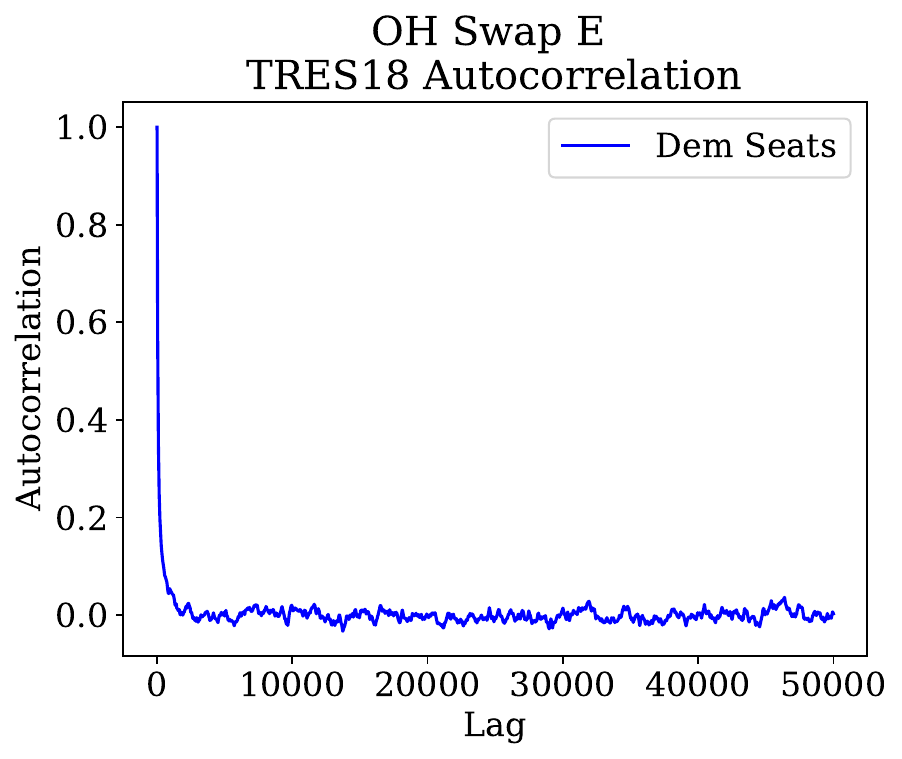}
\includegraphics[scale=.5]{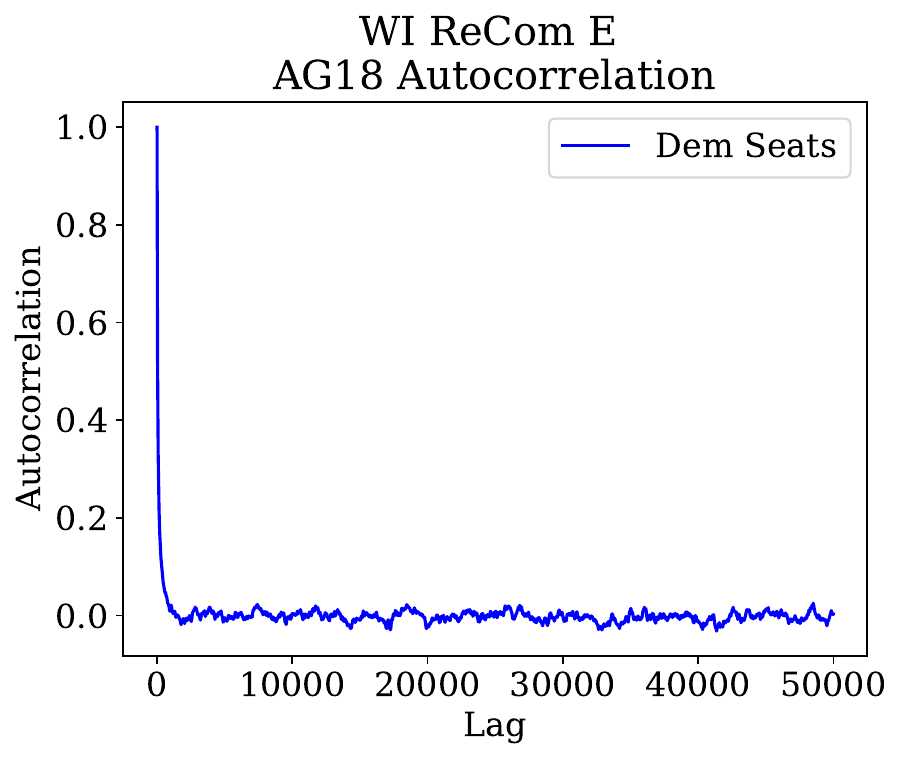}
\caption{Autocorrelation plots for a 3:1 nested Swap ensemble in Ohio and an unnested ReCom ensemble in Wisconsin, started from the enacted Senate map (E) in each state.}
\label{fig: autocorrelation swap}
\end{figure}

\begin{figure}
\center
\includegraphics[scale=.4]{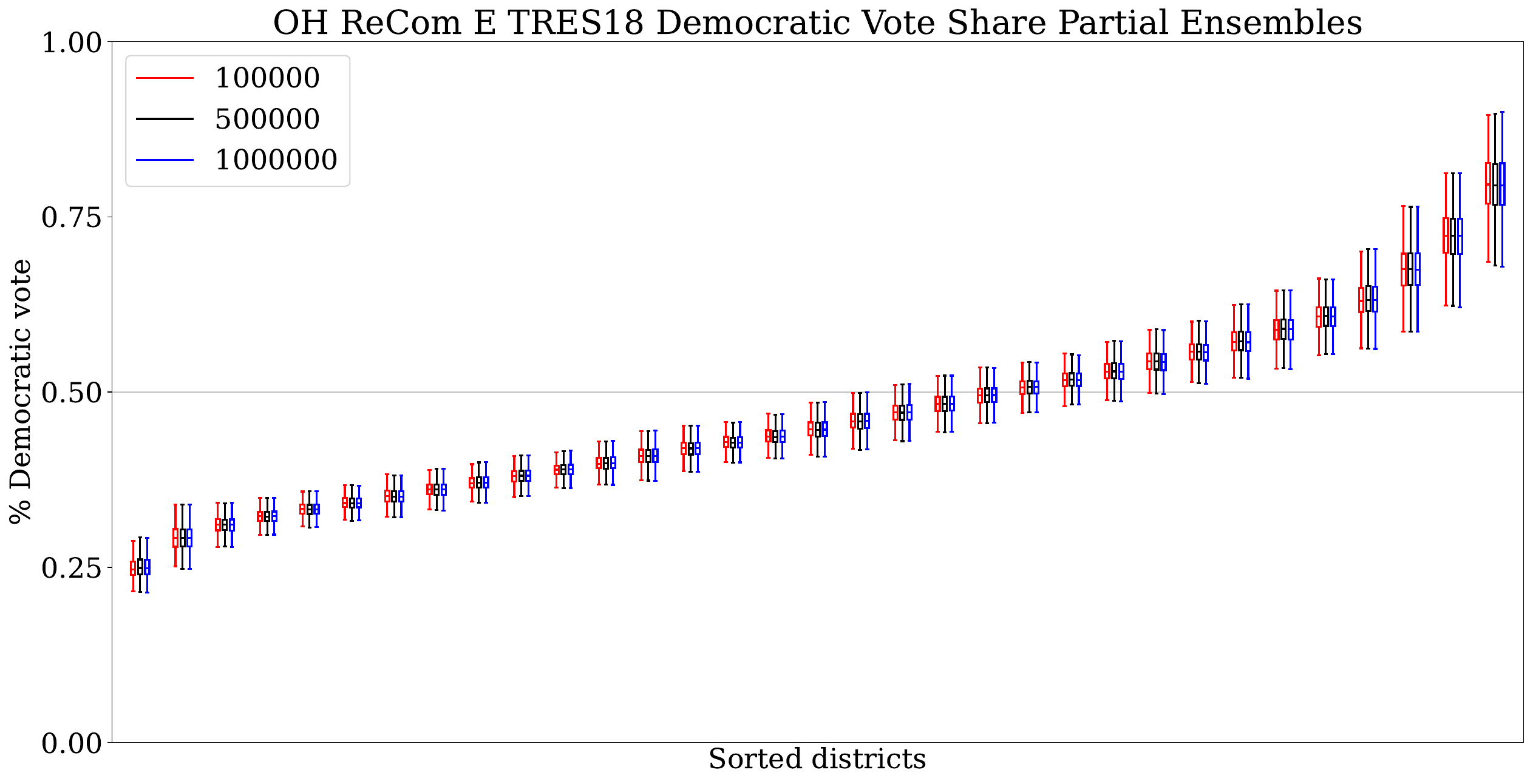}
\includegraphics[scale=.4]{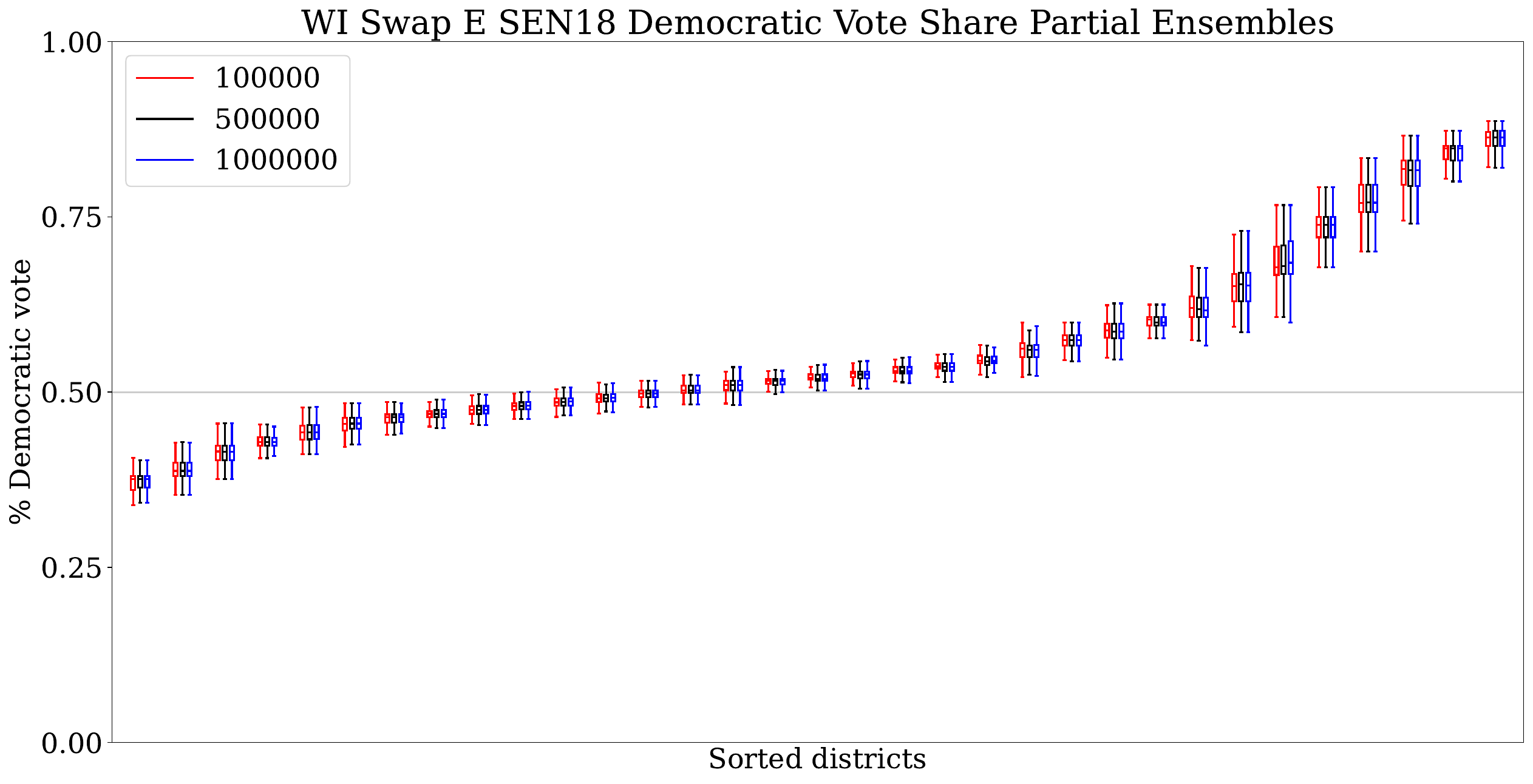}
\caption{Ranked \% Democratic Vote Share in Ohio for 10\%, 50\%, and 100\% of a 1 million step unnested ReCom ensemble and in Wisconsin for a 3:1 nested Swap ensemble, started from the enacted Senate map (E) in each state.}
\label{fig: dif size ensembles swap}
\end{figure}

For both chains, by 2,000 steps,  the autocorrelation for seats won by Democrats begins to hover around 0.
See Figure \ref{fig: autocorrelation swap}.
In both Ohio and Wisconsin, we see that  50\%, and 100\% of a 1 million step ensemble have nearly identical distributions of ranked \% Democratic vote share for each chain type.
See Figure \ref{fig: dif size ensembles swap}.

\begin{figure}
\center
\includegraphics[scale=.45]{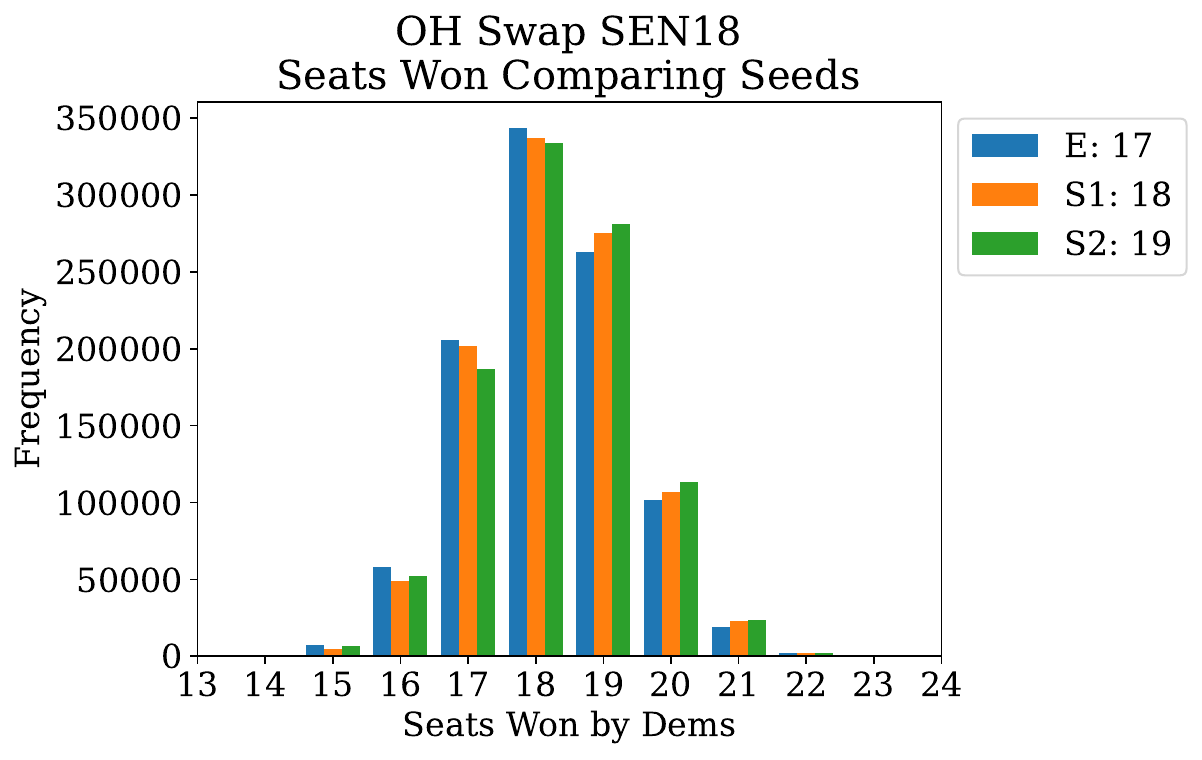}\includegraphics[scale=.45]{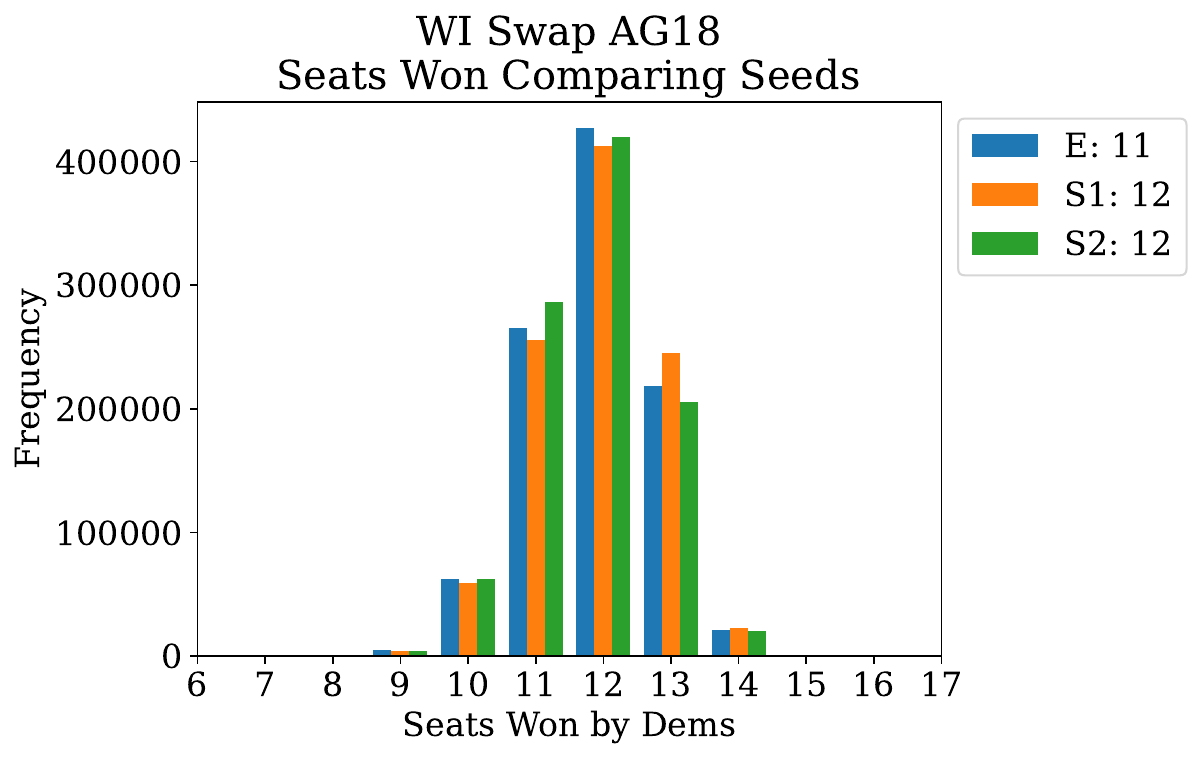}
\caption{The number of seats won by Democrats across different seed plans for Ohio and Wisconsin 3:1 nested Swap ensembles.  The number next to the seed name indicates the number of seats won by Democrats under the seed plan.}
\label{fig: different seeds seats swap}
\end{figure}

In the 3:1 nested setting, a different seed is an alternative Senate map on the same House map.
We make use of a method included in the gerrychain Python package that generates random initial seeds; we call these randomly generated seeds S1 and S2.
In both states, under any election data, we see very similar histograms of seats won regardless of the starting seed.
This is a strong indication of convergence.
See Figure \ref{fig: different seeds seats swap}.

\begin{figure}
\center
\includegraphics[scale=.45]{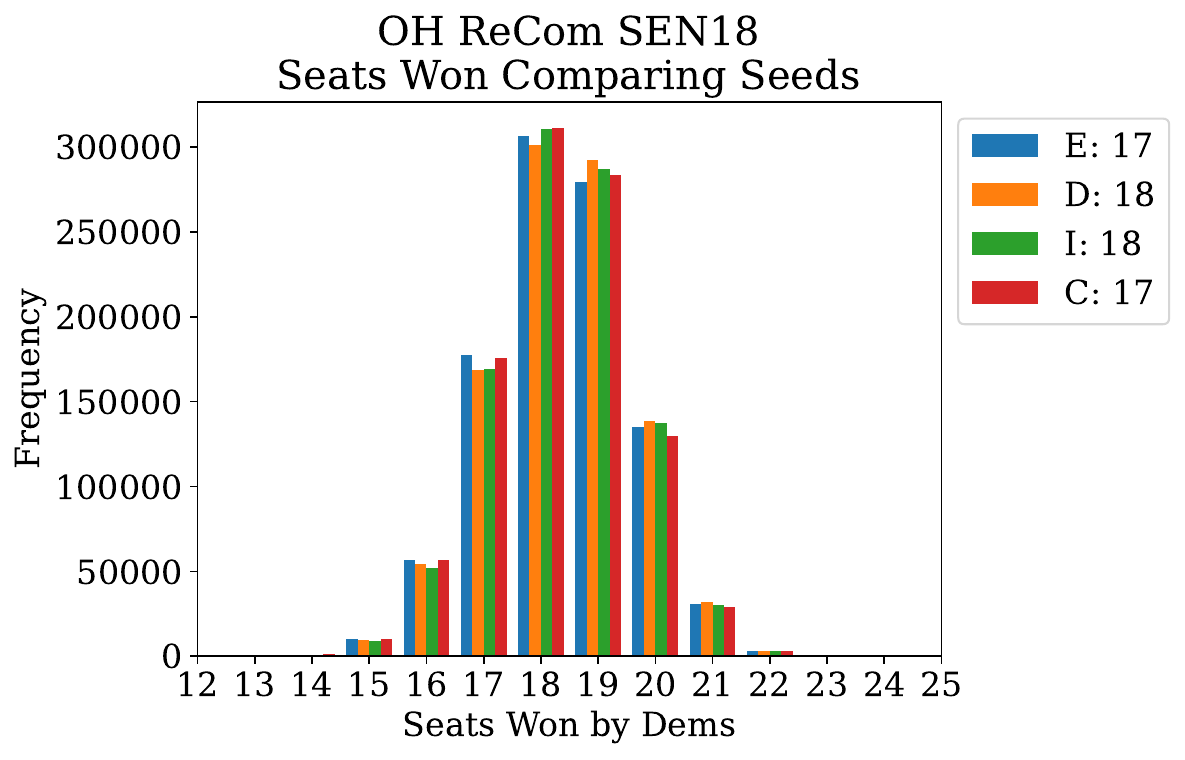}\includegraphics[scale=.45]{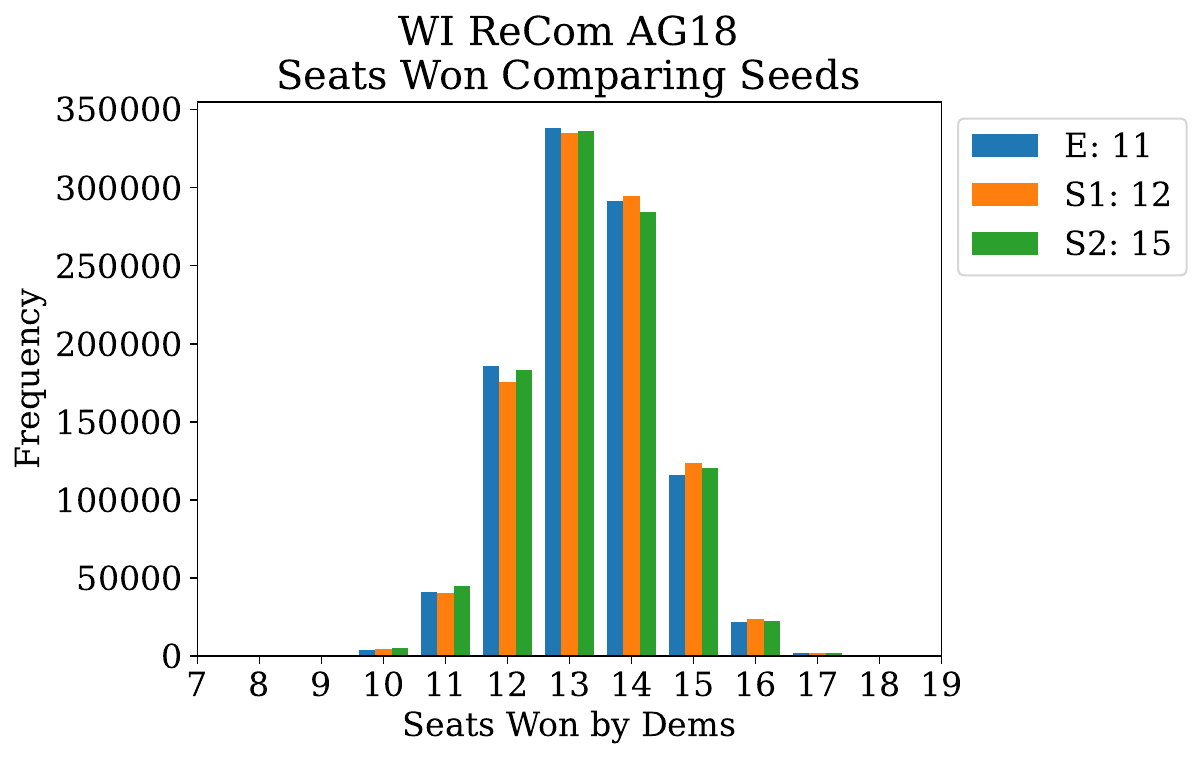}
\caption{The number of seats won by Democrats across different seed plans in unnested  ReCom ensembles.  The number next to the seed name indicates the number of seats won by Democrats under the seed plan.}
\label{fig: different seeds seats recom}
\end{figure}

In the unnested setting, a different seed is just an alternative Senate map on the precincts/wards.
The Ohio Redistricting Commission posts submitted Senate plans in a digital format, thus allowing us to have different seeds for the unnested chain. 
We take the Sykes/Russo Democratic plan (D),  the Johnson McDonald Independent plan (I),  and the Ohio Citizens' Redistricting Commission plan (C) as three alternate seeds for ReCom.
For Wisconsin, alternate Senate plans are not submitted in a format that is easy to digitize, so we again use gerrychain to construct two new seeds for Wisconsin (S1, S2).
In Ohio and Wisconsin, we see that our choice of seed for ReCom also does not impact the distribution of seats won by Democrats.
This is a strong indication of convergence.
See Figure \ref{fig: different seeds seats recom}.

\section{Results}
\subsection{Comparing Ohio Nested and Unnested}
We now compare the behavior of the two ensembles to each other.
In some sense,  we are taking the unnested ensemble to be the ``control" and the nested ensemble tells us how much the 3:1 nesting rule impacts the outcome.

\begin{figure}
\center
\includegraphics[scale=.45]{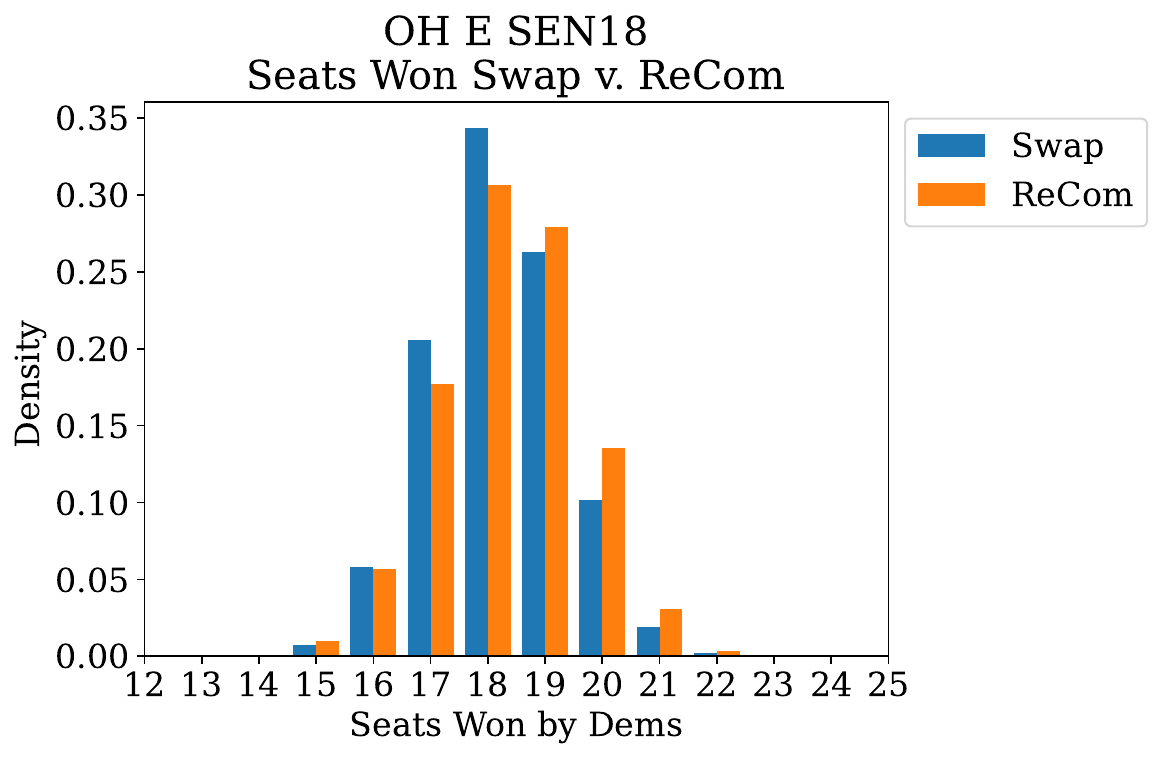}\includegraphics[scale=.45]{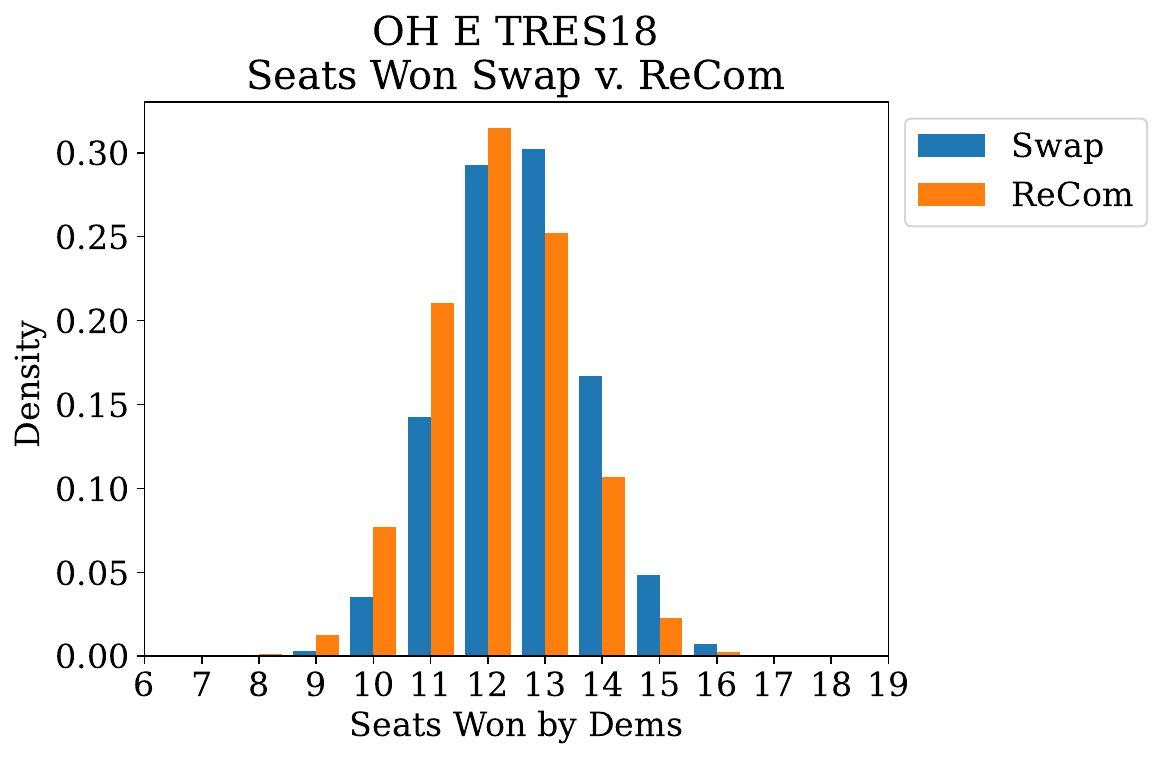}

\caption{Comparing the nested and unnested ensembles in Ohio under SEN18 and TRES18. Chains were started from the enacted Senate map (E).}

\label{fig: ohio recom v swap}
\end{figure}

Under SEN18,  the nested and unnested ensembles behave almost identically.
As seen in Figure \ref{fig: ohio recom v swap},  the only visible difference is that the nesting requirement seems to shift the distribution marginally to the left, in favor of Republicans.
But this shift is negligible in terms of the number of seats;
the Swap ensemble has a range of 14 to 23 seats, while the unnested ReCom ensemble has a range of 13 to 24 seats.
While the nesting requirement does narrow the distribution of seats won by one seat in each direction, the unnested ensemble hardly ever samples the 13 and 24 seat maps.
In other words, the bulk of the distributions is the same.

Under TRES18,  the two ensembles also behave almost identically, except now the nesting requirement shifts the distribution of seats won slightly to the right.
See Figure \ref{fig: ohio recom v swap}.
The Swap ensemble has a range of 8 to 17 seats, while the unnested ReCom ensemble has  a range of 7 to 18 seats.
Again, the range of possible seat values is nearly identical, particularly given how few of the 7 and 18 seat maps are sampled.
Thus, in Ohio, the 3:1 nesting requirement does not significantly impact the number of seats won.

\subsection{Comparing Wisconsin Nested and Unnested}
\begin{figure}
\center
\includegraphics[scale=.45]{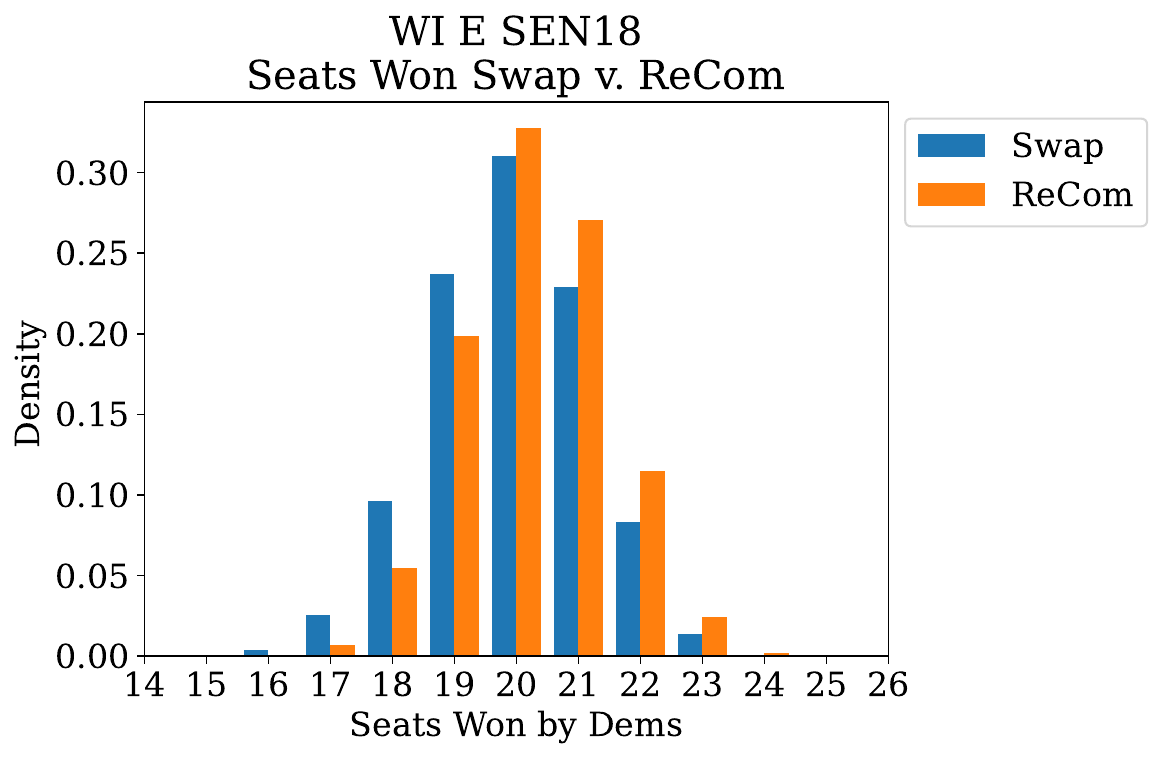}\includegraphics[scale=.45]{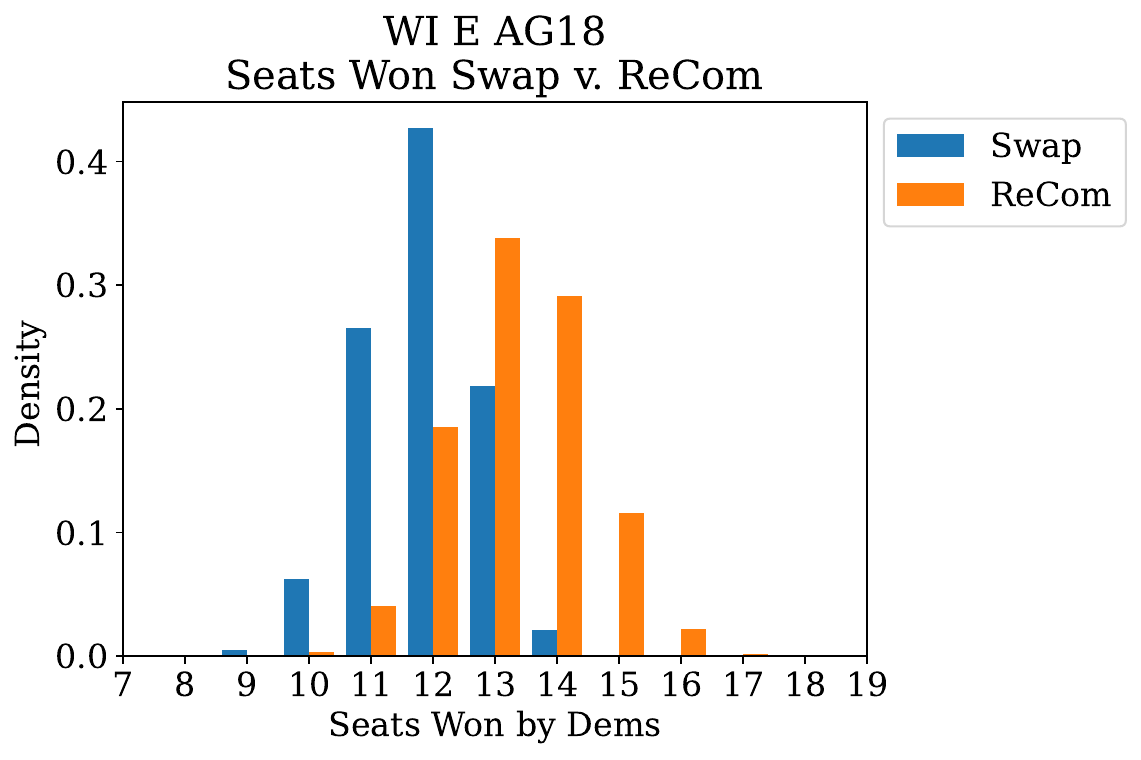}

\caption{Comparing the nested and unnested ensembles in Wisconsin. Chains were started from the enacted Senate map (E).}
\label{fig: wisconsin recom v swap}
\end{figure}

We compare the behavior of the nested and unnested ensembles for Wisconsin.
See Figure \ref{fig: wisconsin recom v swap}.
Under the SEN18 data,  the nested chain is shifted marginally to the left, in favor of the Republicans, but the range of possible seats is nearly identical in both ensembles.
In the Swap ensemble, the range is 15 to 24 seats, while in the ReCom ensemble it is 15 to 25 seats.

The AG18 election data is the only data for which we see a significant change in distribution.
Here, the bulk of the nested ensemble is shifted in favor of Republicans, with the range of seats changing from 9 through 18 (unnested) to 8 through 16 (nested).
Even with the visible difference in the histograms, we still find that the nesting requirement only reduced the maximum number of seats won by 2.
It is interesting to note that in Ohio,  the distributions shifted in either direction, but in Wisconsin, the nesting requirement seems to shift the distributions in favor of the Republicans regardless of the election data.

\section{Biasing the House Map}
It is natural to wonder what impact the underlying House map has on the distribution of possible Senate maps.
In order to study this we used the short burst algorithm to generate House maps with extreme numbers of seats won,  both for Democrats and then again for Republicans \cite{cannon2022voting}.
We then took these biased House maps as our underlying dual graphs, and ran the Swap chain to see how this impacted the distribution of Senate seats.
We also used a neutral map as a control.

\begin{figure}
\center
\includegraphics[scale=.5]{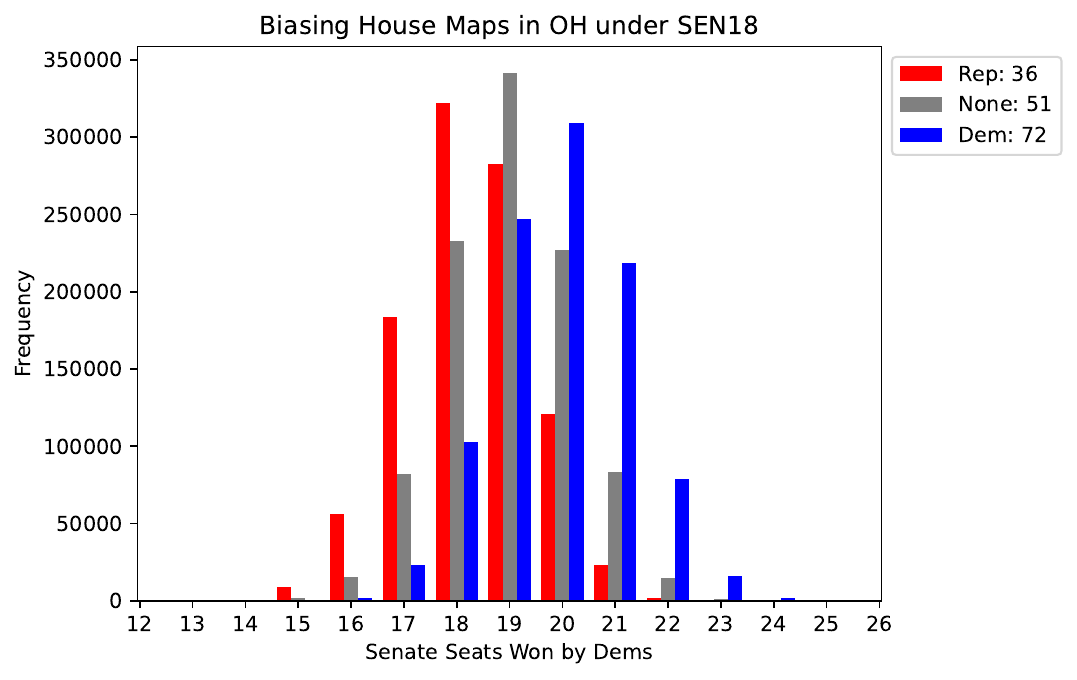}\includegraphics[scale=.5]{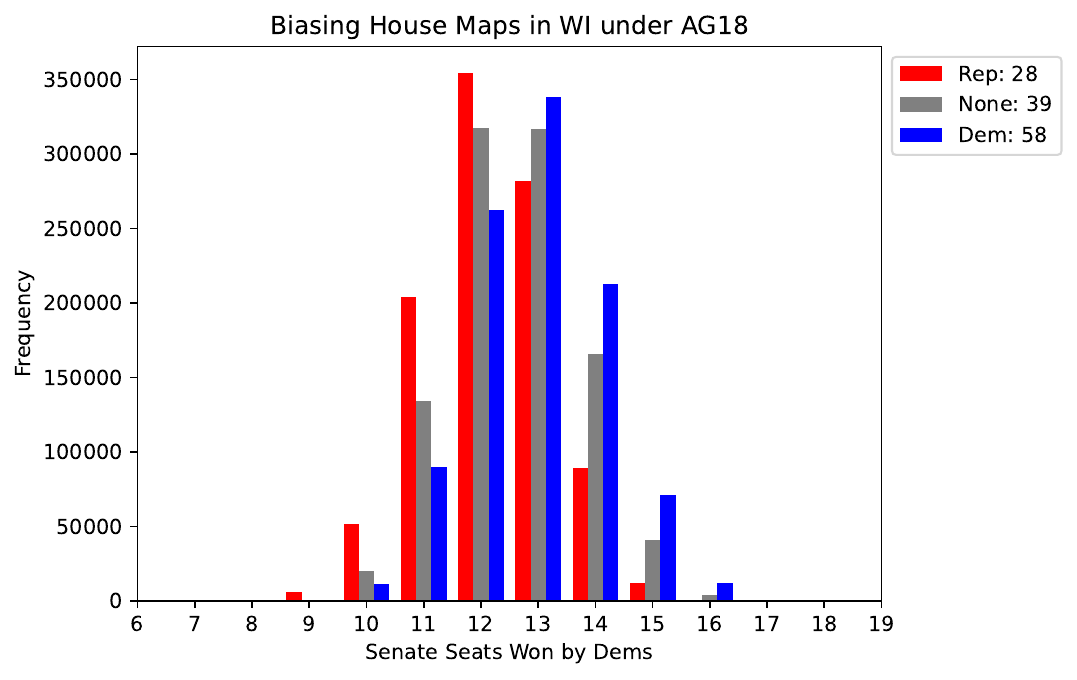}
\caption{Three Senate distributions generated by Swap ensembles based on three different house maps: one with Republican bias, one with Democrat bias, and one without any bias. The number in the legend indicates the number of seats won by the Democrats at the House level under the chosen House dual graph.}
\label{fig:biasing house maps}
\end{figure}

Despite enormous differences in the number of seats won for Democrats at the House level,  the choice of House map had very little impact on the distribution of Senate seats.
See Figure \ref{fig:biasing house maps}.
There was a slight shift in the distributions, in favor of whatever party we were biasing for, but the range of each distribution was almost identical.
The range is really what matters in the context of redistricting; this experiment suggests that regardless of how the House map was drawn,  there is a wide range of possibilities for the Senate map.
This is in some ways expected.
If gerrymandering is a process that ``happens in the margins," the act of 3:1 nesting tends to erase carefully drawn boundary lines at the Senate level, resulting in a more neutral distribution.

\section{Conclusion}
In summary,  with the exception of the Wisconsin AG18 data,  the 3:1 nesting rule seems to have little impact on the distribution of Democrat seats won.
This echoes the results of \cite{nested_districts}. 
Even in the case of AG18, the nesting requirement only reduced the maximum number of seats won by 2.
We also found that biasing the underlying House map in favor of Democrats or Republicans did not significantly change the possible Senate distribution.
This suggests that the 3:1 nesting rule does not necessarily transfer bias from the House level to the Senate level, giving map makers the chance to draw a neutral map regardless of the initial House districts.

Some future directions include:
\begin{enumerate}
\item In order to make more realistic Senate maps,  it would be useful to implement Ohio's county splitting requirements. 
One promising avenue for this is to use Multi-Scale Merge-Split, a recently developed Markov chain designed to preserve hierarchical structures, like counties \cite{autry2020multiscale}.

\item If biasing the House maps for seats did not separate the Senate distributions,  what would cause a separation? 
In other words, what properties of the House map actually impact the possible 3:1 nestings?

\end{enumerate}

\subsection{Acknowledgments}
The author would like to acknowledge Daryl DeFord for helping start this project and finding SHP files for Ohio,  Moon Duchin and Thomas Weighill for helpful comments and suggestions on an earlier draft, and his advisor Matthew Kahle.

\bibliographystyle{plain} 
\bibliography{3_1_nesting.bbl}

\begin{thebibliography}{10}

\bibitem{mggg_grid_partitions}
The known sizes of grid metagraphs.
\newblock \url{https://mggg.org/table.html}.
\newblock Accessed: 2023-03-28.

\bibitem{ballotpedia}
State legislature.
\newblock \url{https://ballotpedia.org/State_Legislative_Districts}.
\newblock Accessed: 2023-03-15.

\bibitem{autry2020multiscale}
Eric~A. Autry, Daniel Carter, Gregory Herschlag, Zach Hunter, and Jonathan~C.
  Mattingly.
\newblock Multi-scale merge-split markov chain monte carlo for redistricting,
  2020.

\bibitem{nested_districts}
Sophia Caldera, Daryl DeFord, Moon Duchin, Samuel~C. Gutekunst, and Cara Nix.
\newblock Mathematics of nested districts: The case of {A}laska, 2020.

\bibitem{cannon2022voting}
Sarah Cannon, Ari Goldbloom-Helzner, Varun Gupta, JN~Matthews, and Bhushan
  Suwal.
\newblock Voting rights, markov chains, and optimization by short bursts, 2022.

\bibitem{cdonnayNesting}
{Christopher Donnay}.
\newblock {3 to 1 Nesting GitHub Repository}.
\newblock \url{https://github.com/cdonnay/nesting_OH_WI}, {Accessed: July 6,
  2023}.

\bibitem{colorado_in_context}
Jeanne Clelland, Haley Colgate, Daryl DeFord, Beth Malmskog, and Flavia
  Sancier-Barbosa.
\newblock Colorado in context: Congressional redistricting and competing
  fairness criteria in {C}olorado, 2020.

\bibitem{recom_markov}
Daryl DeFord, Moon Duchin, and Justin Solomon.
\newblock Recombination: A family of markov chains for redistricting, 2019.

\bibitem{duchin_walch_2022}
Moon Duchin and Olivia Walch.
\newblock {\em Political geometry: Rethinking redistricting in the US with
  math, law, and everything in between}.
\newblock Birkh{\"a}user, 2022.

\bibitem{durham_2020}
Sonali Durham.
\newblock Nesting rules for political redistricting: Methods for sampling
  matchings and triplings.
\newblock Senior Essay, Berkeley College, Yale University, April 2020.

\bibitem{2018ComparisonOD}
Metric Geometry and Gerrymandering Group.
\newblock Comparison of districting plans for the {Virginia} house of
  delegates.
\newblock 2018.

\bibitem{Karp1972}
Richard~M. Karp.
\newblock {\em Reducibility among Combinatorial Problems}, pages 85--103.
\newblock Springer US, Boston, MA, 1972.

\bibitem{NCSLRedistricting}
{National Conference of State Legislatures}.
\newblock {Redistricting and Census}.
\newblock
  \url{https://www.ncsl.org/redistricting-and-census/2010-redistricting-deviation-table},
  2010.
\newblock Accessed: July 6, 2023.

\bibitem{palmer_schneer_deluca_2022}
Maxwell Palmer, Benjamin Schneer, and Kevin DeLuca.
\newblock A partisan solution to partisan gerrymandering: {the} define-combine
  procedure.
\newblock {\em HKS Faculty Research Working Paper Series}, RWP22-012, Aug 2022.

\bibitem{roy2019convergence}
Vivekananda Roy.
\newblock Convergence diagnostics for {Markov chain Monte Carlo}, 2019.

\bibitem{tuckerfoltz2023locked}
Jamie Tucker-Foltz.
\newblock Locked polyomino tilings, 2023.

\bibitem{ReynoldsVSims}
{U.S. Supreme Court}.
\newblock {Reynolds v. Sims 377 U.S. 533}, 1964.

\end{thebibliography}

\end{document}